\documentclass[manuscript]{aastex631}

\usepackage{threeparttable}

\received{22-Dec-2022}
\revised{16-March-2023}
\accepted{16-March-2023}
\submitjournal{ApJL}

\shorttitle{Cold and Hot}
\shortauthors{Petit et al.}
\graphicspath{{./}{figures/}}

\begin{document}

\title{The hot main Kuiper belt size distribution from OSSOS}

\correspondingauthor{Jean-Marc Petit}
\email{Jean-Marc.Petit@normalesup.org}

\author[0000-0003-0407-2266]{Jean-Marc Petit}
\affiliation{Institut UTINAM UMR6213, CNRS, OSU Theta F25000 Besan\c{c}on, France}

\author[0000-0002-0283-2260]{Brett Gladman}
\affiliation{Department of Physics and Astronomy, University of British Columbia, 6224 Agricultural Road, Vancouver, BC V6T 1Z1, Canada}

\author[0000-0001-7032-5255]{J. J. Kavelaars}
\affil{Herzberg Astronomy and Astrophysics Research Centre, National Research Council of Canada, 5071 West Saanich Rd, Victoria, British Columbia V9E 2E7, Canada}
\affil{Department of Physics and Astronomy, University of Victoria, Elliott Building, 3800 Finnerty Rd, Victoria, BC V8P 5C2, Canada}
\affil{Department of Physics and Astronomy, University of British Columbia, 6224 Agricultural Road, Vancouver, BC V6T 1Z1, Canada}



\author[0000-0003-3257-4490]{Michele T. Bannister}
\affiliation{School of Physical and Chemical Sciences --- Te Kura Mat\={u}, University of Canterbury,
Private Bag 4800, Christchurch 8140,
New Zealand}

\author[0000-0003-4143-8589]{Mike Alexandersen} 
\affiliation{Center for Astrophysics $|$ Harvard \& Smithsonian, 60 Garden Street, Cambridge, MA 02138, USA}


\author[0000-0001-8736-236X]{Kathryn Volk}
\affil{Lunar and Planetary Laboratory, The University of Arizona,
1629 E University Blvd, Tucson, AZ 85721, USA}
\affil{Planetary Science Institute, 11700 East Fort Lowell, Suite
106, Tucson, AZ 85719, USA}

\author[0000-0001-7244-6069]{Ying-Tung Chen} 
\affiliation{Institute of Astronomy and Astrophysics, Academia Sinica, No. 1, Sec. 4, Roosevelt Rd., Taipei 10617, Taiwan}

\begin{abstract}
Using the absolute detection calibration and abundant detections of the OSSOS (Outer Solar System Origin Survey) project, we provide population measurements for the main Kuiper Belt.
For absolute magnitude $H_r<8.3$, there are 30,000 non-resonant main-belt objects, 
with twice as many hot-component objects than cold,
and with total mass of 0.014 $M_\Earth$, only 1/7 of which is in the cold 
belt (assuming a cold-object albedo about half that of hot component objects).
We show that transneptunian objects with 
$5.5 < H_r < 8.3$
(rough diameters 400--100~km) 
have indistinguishable absolute magnitude (size) distributions,
regardless of being in the cold classical Kuiper belt (thought to
be primordial) or the `hot' population (believed to be implanted
after having been formed elsewhere).
We discuss how this result was not apparent in previous examinations of the size distribution due 
to the complications of fitting assumed power-law functional forms
to the detections at differing depths.
This shared size distribution is surprising in light of the  common paradigm
that the hot population planetesimals formed in a higher density environment 
much closer to the Sun, 
in an environment that also (probably later) formed larger (dwarf planet and bigger) objects.
If this paradigm is correct,  
our result implies  that planetesimal formation was relatively insensitive to the local 
disk conditions and that the subsequent planet-building process in the hot population
did not modify the shape of the planetesimal size distribution in this 50--300~km range.
\end{abstract}


CC-BY 4.0 (https://creativecommons.org/licenses/by/4.0/)

\section{Introduction} \label{sec:intro}
The size distribution of objects produced at various stages of
the planet formation process is a topic of intense interest
\citep[eg.][]{2002PASP..114..265K,2016ApJ...817..105K,2017ASSL..445..197O,2007ApJ...659L..61S,2011ApJ...728...68S,2016ApJ...818..175S}
.
One must conceptually separate the size distribution of objects
directly built by some planetesimal formation process from those
that are then created by either collisional grinding or accumulation
\citep{2008ssbn.book..293K,2020tnss.book...25M}
.
In the main asteroid belt there has been heavy collisional modification
which has greatly obscured the initial size distribution, although
arguments that many asteroids were `born big' have been made \citep{2007Natur.448.1022J,2009Icar..204..558M}.
The non-saturated cratering record on Pluto/Charon and Arrokoth
\citep{2019ApJ...872L...5G,2019Sci...363..955S,2020Sci...367.3999S}
argues that in the transneptunian region the size distribution
has not been modified by collisional and accretional effects
since the surfaces of these bodies formed. 
If true this would mean that the currently essentially collisionless
environment of the Kuiper Belt
\citep{2021AJ....161..195A,2019ApJ...872L...5G,2020Sci...367.6620M,2004Icar..168..409P}
has persisted for the Solar System's
age and that the size distributions are thus primordial and preserve
the outcome of the planetesimal formation and planet building process.

The recent study of \citet{2021ApJ...920L..28K} used an ensemble of survey samples
\citep{2009AJ....137.4917K,2011AJ....142..131P,2016AJ....152..111A,2017AJ....153..236P,2018ApJS..236...18B},
referred to collectively as OSSOS++, 
to show that the dynamically cold classical Kuiper belt's size distribution follows 
a power-law with an exponential cutoff at the large size end. 
This `exponential taper' shape is compatible with the initial mass function obtained in simulations of
planetesimal formation in a streaming instability scenario
\citep{2017A&A...597A..69S,2019ApJ...885...69L}
and while other low-density formation scenarios exist
\citep[eg.][]{2016ApJ...818..175S},
which  mechanism created planetesimals is irrelevant for this present manuscript.
Several independent facts
\citep{2004Icar..168..409P,2010ApJ...722L.204P,2003ApJ...599L..49T,2017AJ....154..101P,2019ApJS..243...12S,2019ApJ...872L...5G,2021AJ....161..195A,2020Sci...367.6620M}
also hint at an {\it in-situ} formation of the cold belt in a low density environment. 

The region where the cold belt resides, between the 3:2 and 2:1 mean motion resonance with Neptune, also contains resonant Trans-Neptunian Objects (TNOs) and other non-resonant, yet excited TNOs (objects with either large eccentricity $e$ or inclination $i$, or both) forming the hot belt. 
These objects are commonly stated to have been formed in a region closer to the Sun 
and then implanted in the Kuiper belt during late planetary migration, based on dynamical
\citep[reviewed by][]{2018ARA&A..56..137N}
and compositional 
\citep{2019ApJS..243...12S}     
arguments.
The goal of this manuscript is to use the OSSOS++ sample and compare the absolute magnitude distribution of 
the hot belt to that of the cold belt;
we find that the two distributions are extremely similar over the magnitude range where we have 
high accuracy.

In the next section, we present the OSSOS++ sample of hot objects and compare its size distribution to that of the cold population. We extend the OSSOS++ sample with the MPC database to the large size side of the distribution. Next we present some cosmogonic implications of our findings.

\section{The OSSOS++ hot population absolute magnitude distribution} \label{sec:hot_dist}

We use the OSSOS++ sample 
\citep[see][for full details]{2018ApJS..236...18B} 
to determine the absolute magnitude distribution
of the populations of the main classical Kuiper belt. 
The main-belt classicals are made up of the non-resonant, 
non-scattering objects with semimajor axes $39.4~\textrm{au} < a < 47.7~\textrm{au}$
(that is, between the 3:2 and 2:1 resonances, rejecting all resonant
objects in this range).

As explained in \citet{2019AJ....158...49V} and \citet{2021ApJ...920L..28K}, the best single parameter to discriminate between the cold and the hot populations of the main classical Kuiper belt is the free inclination with respect to the $a$-dependent Laplace plane. 
\citet{2022ApJS..259...54H} provide an improved determination of the local Laplace plane
by double-averaging over the two fast angles rather than from the classical first order secular theory; 
the resulting free inclinations are very stable over time and can be 
found\footnote{The main-belt free inclinations can also be found at {\it http://yukunhuang.com}.}
in \citet{2022yCat..22590054H}. 
Given the distribution of the free inclinations shown in Fig.~3 of  \citet{2022ApJS..259...54H}, 
in our manuscript we elect to use $i_{free} < 4.5^\circ$ as an acceptable split  between 
the main-belt cold and hot populations (understanding there will be interlopers at 
some level). 
The {\it cold}  classical belt only exists at $a>42.4$~au \citep{2008ssbn.book...59K}, 
and is heavily concentrated to perihelia $q>39$~au 
\citep{2011AJ....142..131P} and
\citep[see Fig.~5 of][]{2021ARA&A..59..203G}.  
We here adopt these simple cuts to define the cold classical belt region, arguing
that the higher eccentricities for TNOs with $q<39$~au and $a>42.4$~au indicate they
have suffered dynamical excitation.
The {\it hot} main-belt population is then all objects not in the cold sample.
This yields 327 cold and 219 hot objects in OSSOS++.


\subsection{Sample characterization}
\label{sec:charact}

We use the method in \citet{2021ApJ...920L..28K} to debias the orbit and $H_r$ distributions of the OSSOS++ detections\footnote{Divide the phase space in small $(a, q, sin(i), H)$ cells; for any cell with a detection in it, determine the detection biais of this cell using the Survey Simulator.}.
The most difficult TNO orbits to detect, at a given $H_r$, are those with  near-circular 
orbits at the $a \simeq 47$~au outer edge of the main belt. 
Objects on such orbits always remain at distance $d \simeq 47$~au, while those with either 
lower $a$ or larger eccentricity $e$ will spend some time closer to the Sun and be 
visible for some fraction\footnote{Using $H_r=9.0$ as an example, $a=47$~au circular orbits
can never be observed by OSSOS, while $H_r=9$ objects distributed along an $e\simeq$0.1 orbit
rise above the flux threshold close to perihelion and give one sensitivity to that set of $a,q$
orbital elements for the purposes of debiasing (where debiasing is essentially taking into
account Kepler's Second Law).  The method has no ability, however, to debias the orbit for
which there is never any possibility of detection.}
of their orbital longitude.
Because  OSSOS++ reached apparent magnitude $m_r \gtrsim 25$ for some blocks, $H_r \simeq 8.3$ TNOs are 
visible at 47~au;
\citet{2021ApJ...920L..28K} used this $H_r$ for the limit down to which we trust 
our debiasing for the cold population; even if the higher $e$'s of hot TNOs
give a mild increase in sensitivity to $H_r>8.3$ hot-object orbits, we maintain
the limit at 8.3 since our goal is to compare the two populations as a 
function of absolute magnitude.
Because this debiasing method uses only the detections, this model will certainly increasingly 
underestimate TNO numbers beyond these sensitivity limits.



\begin{figure}[htbp]
\centering\includegraphics[width=0.7\linewidth]{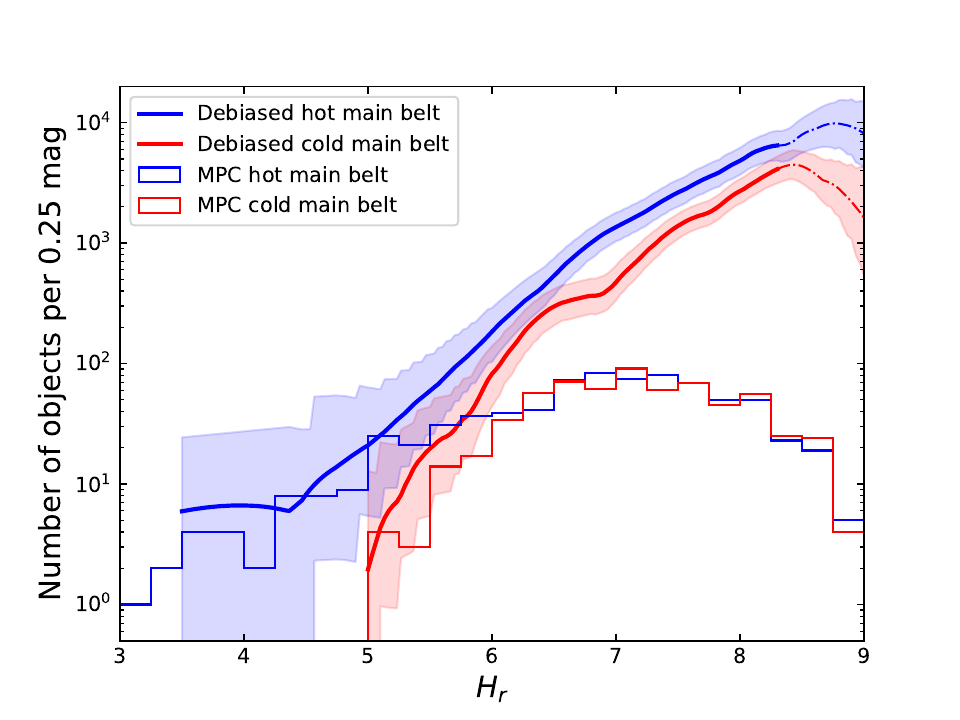}
\caption{Calibrated differential $H_r$ distribution of the main-belt hot (blue bold curve)
  and cold (bold red curve) components 
  from debiasing the OSSOS++ sample for
  $H_r \in [3; 8.3]$. 
  Less secure estimates (see text) for $8.3<H_r<9$, which are only lower limits, 
  are shown by dash-dotted curves. 
  These curves are Epanechnikov Kernel Density Estimates
  (appendix~\ref{app:KDE}) of the debiased $H_r$ absolute magnitude
  distributions of the main classical belt objects in OSSOS++. 
  The dotted lines represent 95\% confidence intervals from Poisson statistics.
  Up to $H_r\simeq5.3$ the OSSOS absolute calibration reproduces the MPC numbers,
  indicating the latter is now essentially complete in the main-belt. 
}
\label{fig:KDE_H_OSSOS}
\end{figure}

Simple binned histograms to represent a differential distribution are subject to large fluctuations 
when the sample is not very numerous.
To avoid this shortcoming, we use a Kernel Density Estimator (KDE) method which spreads out each detection over a kernel of some size; each detection thus contributes to the differential distribution not only at its exact position but on some interval with a varying weight 
(see appendix~\ref{app:KDE}).
Figure~\ref{fig:KDE_H_OSSOS} presents these debiased differential TNO numbers 
(per 0.25 magnitudes, in order to compare with the known sample in the Minor Planet Center, MPC) 
as the hot and cold populations implied by the OSSOS++ detections. 
MPC $H$ magnitudes are converted to $H_r = H_{MPC} - 0.2$; 
this is the known shift between OSSOS++ $H_r$ magnitudes and the $H$ stated in the MPC.   
The rollover in the cold distribution for $H_r>8.3$ where we expected to
insensitivity to begin is clear; 
the rollover in the hot distribution begins a few tenths of a magnitude fainter
due to the ability to detect the lower perihelion objects present in that population
(as mentioned above).

Several important results follow immediately from Fig.~\ref{fig:KDE_H_OSSOS}.
First, the MPC inventory (shown as histograms) of the main belt is very close to complete for
$H_r<5.3$; this was already clear for cold TNOs \citep{2021ApJ...920L..28K} but 
our estimates reinforce the idea that the hot {\it main-belt population} is also 
now essentially  complete.
\citet{2011AJ....142...98S} already suggested that the TNO inventory was nearly complete to apparent 
magnitude $M_R \simeq 21$, corresponding to
$H_r \simeq 6.4$ at 30~au and 
$H_r \simeq 4.2$ at 50~au.
Despite a burst of $H_r<5.3$ discoveries during 2013-2015 by Pan-STARRS \citep{2016arXiv160704895W}
no bright main-belt TNOs have been discovered despite continued operations, signifying
completeness to this magnitude.
The agreement of our estimates with the complete population at the bright
end shows that our debiasing method yields correct number estimates;
these estimates then establish that $H_r>5.8$ is largely incomplete
(the known TNOs are below our 95\% confidence intervals).
The up-coming LSST will survey the main classical Kuiper belt down to $H_r \simeq 7.5$ at 47~au;
Figure~\ref{fig:KDE_H_OSSOS} indicates that LSST will mostly discover TNOs at $H$ magnitudes 
for which completeness in 2022 is still only $\simeq$5\%.
Secondly, one is struck that the hot and cold populations have rather similar $H_r$ distribution 
shape in the range [5.8, 8.3].
This is surprising because many papers
\citep{2004AJ....128.1364B,2005AJ....129.1117E,2011AJ....142..131P,2014AJ....148...55A,2014ApJ...782..100F}
conclude that the hot and cold components have different absolute magnitude 
distributions.
Here we suggest that these differences might be dominantly confined only to the largest
($H_r<6$) TNOs of the populations; 
if true, this has major implications for the planetesimal formation and planet-building process.

\subsection{Comparison of the hot and cold distributions}

\begin{figure}[htbp]
\centering\includegraphics[width=0.7\linewidth]{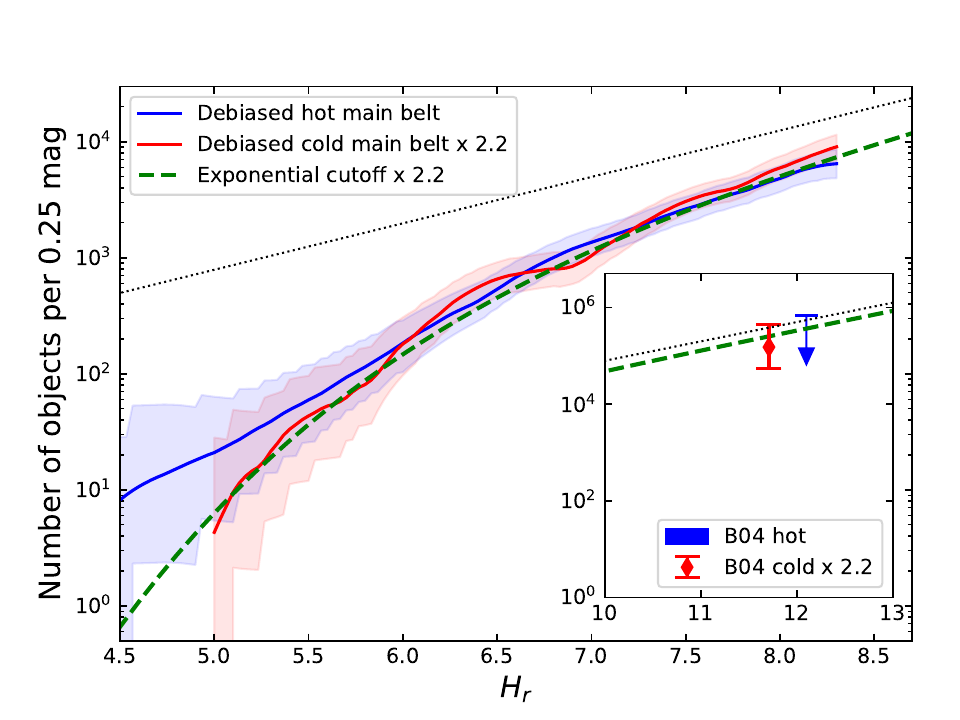}
\caption{Similar to Fig.~\ref{fig:KDE_H_OSSOS}, 
where here the debiased cold component estimates and uncertainties have been scaled 
upward by a factor of 2.2.
The green dashed line shows the
\citet{2021ApJ...920L..28K}
cold population exponential cutoff fit (with large-$H$ asymptotic slope of 0.4),
also scaled upward  by a factor of 2.2.
The inset shows the continuation of the scaled exponential cutoff fit at larger/fainter $H_r$;
the red error-bar gives the 95\% confidence range from the detection of three 
cold TNOs using HST by
\citet{2004AJ....128.1364B}, and is compatible with the above extrapolation.
The inset's blue arrow shows the 95\% confidence upper limit from the
non-detection of hot population TNOs in that same HST study.
For $H_r>5.5$ the shape of the hot and cold population's are surprisingly similar, which hints at similar formation processes. The black dotted line is a reference exponential with logarithmic slope $\alpha = 0.4$, to help see how the exponential taper deviates from a single exponential.
}
\label{fig:KDE_H_shift}
\end{figure}

Figure~\ref{fig:KDE_H_shift} shows a clearer representation of the shape similarity of the two populations.
Note that OSSOS++ has a sufficiently large number of detections that construction of a differential $H$-magnitude and good resolution is possible; many past analyses have shown cumulative distributions. 
Here, we have multiplied the number of cold TNOs by a factor of 2.2, which matches the
curves at $H_r=6$.
Given the uncertainties shown by the 95\% confidence ranges and the expectation to have a small
part of this range at $\simeq2 \sigma$ discrepancy, the 
shapes are nearly identical
in the range from 5.5$<H_r<$8.3, with the most compelling indication of a
difference at $H_r<6$ where the exponential taper cuts off
the cold population.
For $H_r<5$, Fig.~\ref{fig:KDE_H_OSSOS} makes it clear that the hot population contains
large objects while the cold belt has none.

A $\chi^2$ test can quantitatively evaluate if one can reject the null hypothesis 
that the two observed  distributions are drawn from the same underlying distribution 
in the range from $\simeq$5.5--8.3.  
In order to mitigate binning effects, we varied differential bin sizes (0.2, 0.3, or 0.4 mag) 
and the histogram starting magnitude (by steps of 0.1 mag), 
yielding probabilities that the two distributions are from the same underlying distribution 
ranging from 55\% to 90\%; 
the idea that the distributions are identical is clearly plausible.
Fig.~\ref{fig:KDE_H_shift} also shows the exponential taper function derived in  
\citet{2021ApJ...920L..28K};   
another $\chi^2$ test to compare the hot $H_r$ distribution to that function 
(only shifting the function to match the debiased hot number at the single
value of $H_r = 8.0$)
yields probabilities  between 50\% and 80\%
of drawing the hot population from the same functional form as the cold
in this same $H_r$ range.

\citet{2021ApJ...920L..28K} showed that if one extends the $H_r$-distribution beyond 8.3 with a
functional form asymptotic to an exponential law $dN/dH \propto 10^{0.4H}$, the resulting 
estimated differential numbers are consistent with the detection of three $H_r<12$ cold
TNOs in the HST search of \citet{2004AJ....128.1364B}.
A similar extrapolation of our hot main-belt population estimate is consistent
with no hot detections in the HST search.\footnote{Despite the hot population having
a factor of two more TNOs than the cold population at each $H_r>5.8$ magnitude, non-detection
in the HST survey is understood when one considers that these hot objects are spread out over
an order of magnitude more sky area due to their larger orbital inclinations.}
An asymptotic power of $0.4H_r$ is supported for $H_r\simeq12-17$ by the crater record
on Charon \citep{2019Sci...363..955S}, which has been estimated to be dominated by hot population
projectiles \citep{2015Icar..258..267G}.

Thus, there is as yet no firm evidence
that the shape of the $H$ magnitude distributions of the hot and 
cold are different for $H_r \gtrsim 5.5$.
But if the cold classical belt is formed in situ in a low-density environment
(exhibiting an exponential cutoff at a size scale set by local conditions at $\simeq 44$~au)
and the hot object implanted after being formed much closer to the Sun (having a 
different formation, collision, and dynamical history), one might reasonably
expect them to have very different $H$ magnitude distributions.
We will return to this issue in the discussion section after discussing the 
small-$H$ regime, where the two populations differ markedly.

\subsection{The large size tail}

The sample of large bodies from the main classical Kuiper belt in the MPC database is very close to complete for the cold \citep[][and 
Figure~\ref{fig:KDE_H_OSSOS}]{2021ApJ...920L..28K} and the hot (Figure~\ref{fig:KDE_H_OSSOS}) populations down to $H_r \simeq 5.3$. 
Hence the $H_r$ distribution at large sizes can be directly obtained from the MPC database.
The debiased OSSOS++ sample fails to explore the  $H_r < 4$ range due to limited sky area covered and the resulting tiny number of 
detections (only one hot main-belt object brighter than $H_r = 5$ and three with $H_r < 5.5$).

\begin{figure}[htbp]
\centering\includegraphics[width=0.7\linewidth]{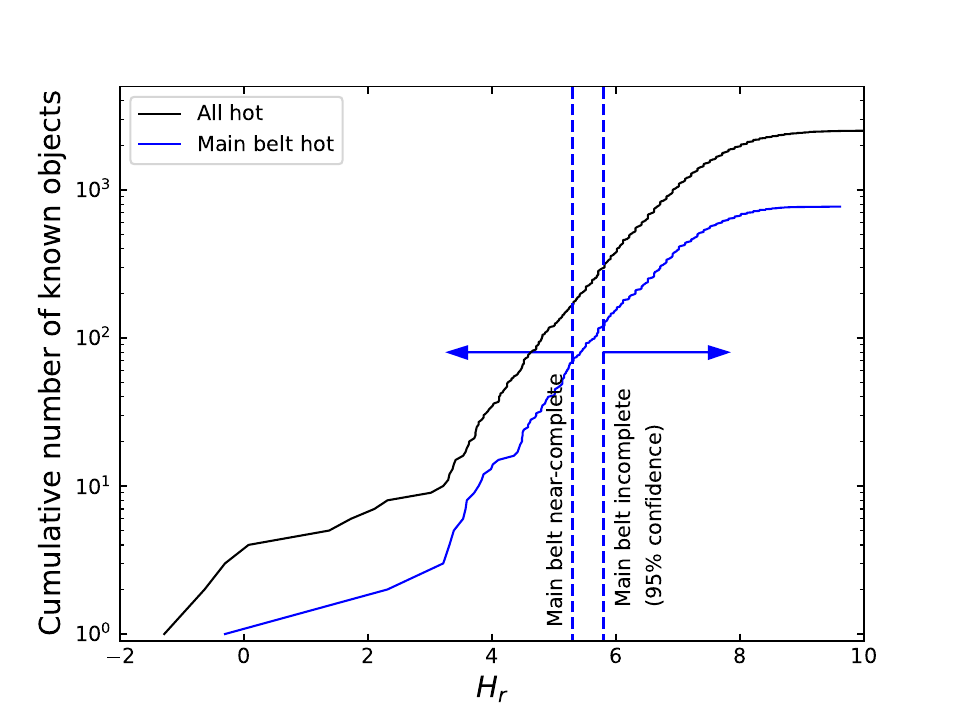}
\caption{Cumulative $H_r$ distribution (black curve) of all hot TNOs
(here, $a > 29.5$~au and not in the cold population) in the MPC database as of 
13 January 2022. 
The blue curve shows the main-belt subset.
The vertical line at $H_r = 5.3$ represents where our analysis indicates that the the hot main-belt population
is nearly complete. 
The vertical line at $H_r = 5.8$ shows where the MPC sample is certainly incomplete (95\% confidence level).
The two samples share very similar $H_r$ distributions brighter than $H_r = 5.3$ 
(in particular both distributions show an obviously shallower slope for $H_r<3$),
supporting the  claim that all hot objects have the same origin. 
}
\label{fig:cum_H_MPC_bright}
\end{figure}

Figure~\ref{fig:cum_H_MPC_bright} shows the cumulative number of objects present in the MPC database 
for hot main classical belt (blue line) and all hot
(black line) objects. 
It is noteworthy that the full hot sample, although obviously incomplete (because even the largest objects 
cannot be seen at the large distances reached by the scattering and resonant populations), is parallel to the hot 
main-belt sample, which is complete bright-ward of $H_r \simeq 5.3$.
The abrupt change at $H_r\simeq 3.3$ in both distributions has been evident \citep{2008ssbn.book..335B} 
since large shallow surveys first covered most of the sky and continues to signal an abrupt change of 
exponential index between 0.14 and 0.6 at this magnitude \citep{2021Icar..35613793A}.
We will refer to the $H_r < 3.3$ range as the `dwarf planet' regime\footnote{Although 
the current IAU dwarf planet definition involves difficult to know properties related to `roundness' 
\citep[e.g.][]{2008Icar..195..851T,2019Icar..334...30G},     
we point out that using a simple $H_V<3.5$  definition would make the
terminology adapt to some obvious transition in the object distribution.}
below.

Fortuitously, the bright end of OSSOS++ happens to lie at the hot main-belt completeness limit.
The argument above indicates we can graft our debiased hot distribution onto the MPC sample and,
with the parallel distribution the full hot sample allows one to have access to the $H_r$-magnitude 
distribution from Pluto/Eris scale down to $H_r=8.3$
(and further by reasonable extension to the HST study's depth at $H\simeq12$).
The hot-cold comparison
indicates that consistency of the $H_r$-magnitude shape for $H_r \gtrsim 5.5$ is
 not also true at bright absolute magnitudes.
The cold population exhibits the exponential cutoff while the hot population (both in the
main belt and in general) not only lacks the exponential cutoff but has very 
shallow distribution as one goes up in size into the dwarf planet regime.
We will interpret this in the Discussion section below. 

\section{Consideration on past surveys}
\label{sec:past}

Because of the curved shape of the exponential cutoff, representing the
$H$-distribution by a single exponential $N(< H_r) \propto 10^{\alpha H_r}$ yields different logarithmic slopes $\alpha$ depending
on the area and depth of the survey. A shallow, wide survey will probe the steep part
of the curve, while a fainter, narrower survey will probe a shallower part of the curve.
The case of the hot component is even worse, due to the change between the logarithmic slopes of
$\sim$0.6 in the range $H \in [3.5; 6]$ and 0.14 for brighter objects
\citep{2021Icar..35613793A}. A single exponential representation will thus require a
very shallow slope for wide, shallow survey, to medium steep for intermediate surveys,
and then again again shallow slopes for fainter surveys.
Many of the early works did not separate between the hot and cold populations.
Surveys close to the ecliptic were dominated by the cold component, while surveys
mostly out of the ecliptic missed much or all of the cold population.
To blur things even more, these studies used the apparent magnitude, thus convolving the $H$-distribution with the distance distribution \citep[see][for a review]{2008ssbn.book...71P}.

More recent studies directly determined the $H$ distribution, using either a single
exponential when the range of $H$ magnitudes is small 
\citep[e.g.][]{2011AJ....142..131P,2014AJ....148...55A} or a double exponential when 
the range of $H$ magnitudes was large \citep{2014ApJ...782..100F}.
\citet{2011AJ....142..131P} separately modeled the cold and hot components to their limit of  $H_r \lesssim 7.5$ and obtained different size distributions for the two components,
with population measurements to that $H$-magnitude limit.
The number of $D>100$~km TNOs in the hot and cold main belt given in \citet{2011AJ....142..131P} were extrapolations with these single slopes (0.8 for hot and 1.2 for cold), beyond the range where they were measured. 
It is now clear that these slopes do not represent 
the $H$-magnitude distribution beyond 7.5 (which flattens steadily) and thus the extrapolation overestimated 
the $D>100$~km populations\added{,
likely more so for the cold population than for the hot one}.
When including small (faint) TNO surveys, \citet{2014ApJ...782..100F}
used a double exponential with breaks around $H_r$=7--8
where the bright part was essentially the same single exponentials for the
two components as \citet{2011AJ....142..131P}.

The sparseness of data from homogeneous characterized surveys and the use of particular
functional forms prevented the recognition of the similarity of the hot and cold
size distributions in the few 10s~km to $\sim$300~km range.



\section{Population and mass estimates} \label{sec:pop_mass}

Using our derived $H_r$ distribution, we can provide an accurate debiased 
main-belt  population (which does not include resonant or scattering TNOs) 
as a function of $H_r$ and also number and mass estimates (which are more
uncertain) for TNOs larger than a certain diameter.

\subsection{Population estimates}
\label{sec:pop}

Table~\ref{tab:pop_estimate} gives estimated TNO numbers at the faintest absolute 
magnitude we confidently debias ($H_r \le 8.3$)
and for $H_r$ values corresponding to a diameter $D$=100~km for various 
commonly-used values of the albedo 
(eq.~\ref{eq:final_H_r}).
We added a column with $\nu_r = 0.24$, the albedo of Arrokoth
determined by \citet{2021Icar..35613723H} in $r$ band filter.
Note that here all estimates are based on direct debiasing, which is very uncertain 
for $H_r = 8.9$.
Alternately,
to estimate numbers of $H_r<8.9$ TNOs, one could use the exponential cutoff formula 
(eq.~\ref{eqn:LF_cold}) for the cold population and multiply by 2.2 for the hot population;
Table~\ref{tab:pop_estimate}'s last column would then read 21, 47, and 68 thousand TNOs,
which is identical given the uncertainties.
This is consistent with our previous work at the older measurement limit: \citet[][Table~5]{2011AJ....142..131P}  estimated $N(H_g<8.0) = (8\pm2)\times10^3$, in agreement with our current $N(H_r<7.3) = 6.7^{+1.3}_{-1.6}\times10^3$ if one uses an averaged $g-r\simeq0.7$ color.

\begin{table}[htbp]
\caption{OSSOS++ population estimates from direct debiasing for main-belt non-resonant TNOs ({\bf in $10^3$ objects}). 
Population estimates marked with a * are likely underestimates 
(Sec.~\ref{sec:charact})
because $D=100$~km corresponds to $H_r>8.3$ for $\nu_r=0.04$.  
\added{The 2 left(right) columns in $D \ge 100$~km part 
mostly cover the sample mean for the cold (hot)
populations (respectively);
although the literature gives different average albedos for the hot and cold populations, 
the uncertainty on individual measurements, possible systematics due to modeling, and the sample scatter
means that individual object albedos could always 
cover the whole range, resulting in order of magnitude
uncertainty in population estimates larger than $D$.
In contrast, the $H_r<8.3$ estimates are much more precise.
}
}
\label{tab:pop_estimate}
\centering
\begin{tabular}{c|c||rrrr}
\hline\hline
       $\;$ & $H_r \le 8.3$ & \multicolumn{4}{c}{$D \ge 100$~km}\\
       $\;$ &      $\;$     & $\nu_r=0.24$ & $\nu_r=0.15$ & $\nu_r=0.08$ & $\nu_r=0.04$\\
 component  &      $\;$     & ($H_r\le 7.1$) & ($H_r\le 7.5$) & ($H_r\le 8.2$) & ($H_r\le 8.9$)\\
\hline
Cold & $11 \pm 1$ & $1.2 \pm 0.3$ & $2.8 \pm 0.5$ & $ 9 \pm 1$ & *$20 \pm 2$\\
Hot & $20 \pm 3$ & $2.4 \pm 0.7$ & $ 7 \pm 1$ & $18 \pm 3$ & *$46 \pm 7$\\
\hline
All & $31 \pm 4$ & $3.6 \pm 1$ & $10 \pm 2$ & $27 \pm 4$ & *$66 \pm 9$\\
\hline
\end{tabular}
\end{table}

Our debiasing gives an  estimate of the total population of main-belt
classical TNOs which we can compare to other well sampled small-body
populations.
Out of curiosity, we note that the main classical Kuiper belt contains 340 and 31,000 objects brighter than $H_r = 6.0$ and $8.3$ respectively, while there are only 10 and 145 Main Belt Asteroids (MBAs) brighter than those 
magnitudes. 
Comparison with the jovian Trojan population has more cosmogonic interest because the hot classicals and other excited TNOs 
have been suggested to come from the same primordial population as the jovian Trojans \citep{2005Natur.435..462M}. 
In the $H_r$=7.3--8.3 range, our debiasing shows that the jovian Trojans have a similar 
$H_r$ distribution to the hot TNO population, with the MPC Trojan sample being complete in that range \citep[see][for completeness limits]{2020PSJ.....1...75H}.
There are 8 Jupiter Trojans with $H_r \le 8.3$, and 20,000 hot main classical TNOs, 
thus implantation models of jovian Trojans and hot main classical TNOs must account for 
an efficiency ratio of order $3\times10^3$. 
\citet{2016ApJ...825...94N} and \citet{2016ApJ...827L..35N} state implantation efficiencies of 
$(7\pm 3) \times 10^{-4}$ and $(7.0 \pm 0.7) \times 10^{-7}$ respectively for hot classicals and jovian Trojans;
this model-based factor of 1000 (with an uncertainty of at least a factor of 2) is thus similar to 
the observed factor of 3000.
With OSSOS++, the $H_r<8.3$ hot main-belt population is now the least fractionally uncertain number in this chain of 
reasoning 
but, as Table 1 shows, an order of magnitude population
uncertainty appears due to the albedo uncertainty.

Using the CFEPS population estimates 
  \citep{2011AJ....142..131P,2012AJ....144...23G}
extrapolated to 100~km size,
  \citet{2015Icar..258..267G,2019ApJ...872L...5G} and 
  \citet{2021AJ....161..195A}
  determined the expected number of craters on the surface of 
  Pluto, Charon, and Arrokoth formed in the last 4~Gyr of bombardment.
  They extrapolated the 100~km   
  population down to kilometer-size projectiles using an exponential of slope
  $\alpha = 0.4$, concluding that the recorded crater numbers could be
  produced without a contribution of an early phase (see below) and that the dominant source of craters on Arrokoth is from cold population projectiles.

Our current estimates, however, indicate fewer 100~km bodies than
  the CFEPS extrapolation down to this size, by a factor of 3--30 depending on which albedo is used, which determines $H$ for 100-km
  bodies.)\footnote{\citet{2011AJ....142..131P} used an albedo
  $\nu_g = 0.05$ which means $\nu_r \sim 0.08$ for cold objects (assuming
  $\langle g - r\rangle \sim 0.9$) and $\nu_r \sim 0.06$ for hot objects
  (for $\langle g - r\rangle \sim 0.6$).
  If one instead uses $\nu_r = 0.15$ for cold and $\nu_r = 0.08$ for hot (see Appendix~\ref{app:Mass_belt}), this means a brighter $H_r$ magnitude for 100~km, again decreasing again the number of objects at that size.}
  Starting from $H_\sim6$ down to the limit of our survey, we find that that the
  hot population is 2.2 times larger than the cold.
  For smaller sizes (larger $H$), we {\it assume} that the exponential cutoff
  continues, thus the hot population remains twice the cold at any given $H_r$.
  Eq.~(\ref{eqn:LF_cold}) then yields similar numbers of projectiles at $H_r \sim 17$
  as found by \citet{2015Icar..258..267G,2019ApJ...872L...5G} and
  \citet{2021AJ....161..195A}, with now twice as many main-belt hot TNOs as cold,
  while before the extrapolation produced 2 to 3 times as many cold objects
  as hot due to the incorrect continuation of a very steep slope for cold objects to
  $H_g$=8. The ratio hot-to-cold could be even larger at a specific size (km) when one accounts for the possibly larger albedo for cold objects ($\nu_r \sim 0.15$)
  than for hot objects ($\nu_r \sim 0.08$).
  Remember that all this is an assumption, not directly based on observational
  evidence, and should be taken with a pinch of salt.

Thus all workers should now re-evaluate the Arrokoth cratering rate.
  Pluto and Charon will continue to be dominated by hot-population
  projectiles, while Arrokoth may cease to be dominated by cold-population
  impactors.
  \citet{2021Icar..35614256M} concluded dominance of the hot population
  for Arrokoth crater formation, but their numbers of 100~km
  and 2~km projectiles are not in line with our current estimates and
  should be re-visited.

\subsection{Mass estimates}
\label{sec:mass}

In appendix~\ref{app:Mass_belt},
we estimate the main-belt mass between the 3:2 and 2:1 mean-motion resonances,
integrated over {\it all} sizes.
We use bulk density $\rho = 1000$~kg/m$^{3}$ and albedos $\nu_{r,c} = 0.15$ and
$\nu_{r,h} = 0.08$ for cold and hot TNOs, respectively
\citep{2014ApJ...782..100F,2014ApJ...793L...2L}\footnote{Note that
\citet{2014A&A...564A..35V} reported very similar values of $\nu_{v,c} \sim 0.14$ and $\nu_{v,h} \sim 0.085$ in band $v$ from Herschel and Spitzer observations.}, except for objects larger than
$\sim$500~km, where both density and albedo are known to increase (see Appendices~\ref{app:Mass_hot} and \ref{app:Mass_cold}.

Understanding that masses are uncertain to a factor of 3 due to poorly-constrained 
albedos and densities, we find a cold-belt mass of  0.002~$M_\Earth$ for an 
exponential cutoff shape of $H_r$ and  0.012~$M_\Earth$ for the hot belt, 
for a total mass of the main classical belt of 
0.014~$M_\Earth$.
We find the same hot-belt mass as \citet{2014ApJ...782..100F}, 
but a cold-belt mass that is 7 times larger.
\citet{2001AJ....122.1051G} estimated a total mass for the 30--50~au
distance range  of 0.04-0.1~$M_\Earth$,  using an albedo $\nu_r = 0.04$ 
for all TNOs; 
this includes all dynamical components, 
but because the main belt represents only about half the mass
\citep[determined using the CFEPS model,][]{2011AJ....142..131P} 
in the 30--50~au distance range, a 0.02-0.05~$M_\Earth$ estimate
results.
Using this 4\% albedo, our current approach gives a slightly higher total 
classical main-belt mass 0.06~$M_\Earth$. 

Based on creation of a planetary ephemeris, \citet{2020A&A...640A...7D} estimated
the total Kuiper Belt mass to be $(0.061 \pm 0.001) M_\Earth$, with
unknown model-based uncertainty. 
This is $\simeq$4 times higher than our estimate of the classical belt,
but again only half of this $0.06 M_\Earth$ would be main-belt TNOs.
With the same method, but using a slightly different dataset,
\citet{2018CeMDA.130...57P} estimated the mass to be $(0.02 \pm 0.004)
M_\Earth$, which when restricted to the main belt is in agreement with our 
estimate.

Because the tapered exponential lacks an abrupt break between a 
steep slope power-law
(steeper than $\alpha = 0.6$ for $H_r$ distribution or than $q = 4$ 
for $D$ distribution\footnote{The relations $N(<H) \propto 10^{\alpha H}$ and
$dN/dD(D) = n(D) \propto D^{-q}$ are related by 
$q = 5\alpha + 1$ for a fixed albedo $\nu$.})
to a shallow slope for small sizes, there is less of a concentration
of mass to the typical diameter scale where the break occurs.
For our estimated hot belt, the 25th and 75th  percentile for 
mass are $H_r < 6.7$ and $H_r < 9.8$. 
For the cold population, the 25th percentile is $H_r < 7.3$ and the 
75th percentile is at $H_r < 10.1$. 
Broadly speaking, our precise knowledge of the size distribution includes  
an $H$ magnitude range that contains about half the mass of the 
classical Kuiper belt.

Note that because of the very shallow $\alpha=0.14$ 
\citep{2021Icar..35613793A} in the $H_r<3.5$ hot-object tail,
it is possible that the largest TNOs in this tail were
planetary scale and contained most of the mass of
the hot-population formation region.
These very massive objects have not been retained the hot main-belt 
or the hot population in general due to the $\sim 10^{-3}$ retention 
efficiency of the  scattering out process
\cite{2016ApJ...825...94N},
and thus one should remember
that much of mass in the hot-formation region may have
been sequestered into very large (now absent) planetary
scale objects.

\section{Discussion of cosmogonic implications} \label{sec:cosmo}

Based on several lines of evidence 
\citep[see][and reference therein]{2021ApJ...920L..28K}, 
the cold population is primordial,
meaning its $H_r$ distribution has not evolved since the formation epoch.
Given the shape similarity of the hot and cold populations in the range
$H_r \simeq 5.5$ to 8.3, we hypothesize that the hot population has also preserved its primordial
shape in that size range and the same physical mechanism was responsible for the accretion of
bodies of that size in both the hot and cold forming regions.
The physical conditions ({\it e.g.,} temperature, dynamical time scales) were likely very different 
in these two regions, and in particular the surface density in solids would likely have been
orders of magnitude larger in the hot-forming region than at $\simeq 43$~au \citep{2004Icar..170..492G,2020AJ....160...46N}.
If true, it follows that the formed planetesimal size distribution is 
at most a weak
function 
of the local conditions, at least in the $H_r$=5.5--8.3 range;
over this range it is steeper than collisional equilibrium.
Coupled with the match to the cold
(believed to be primordial and unevolved) population, we thus conclude that at this size scale there was no appreciable collisional modification of the shape.
This implies that the initial phase during which the hot population was in a
dense collisional environment was of shorter duration than the collisional lifetime of $D \simeq 100$~km bodies.
\citet{2022MNRAS.514.4876B} supports this picture, showing that the $D \gtrsim 100$~km size distribution's shape does not change. 
They also conclude that long (100 Myr) instability phases 
do not produce successful matches to observational constraints.

\citet{2020AJ....160...46N} 
and 
\citet{2021ApJ...920L..28K} 
suggested that 
a candidate for this initial planetesimal forming phase is the Gravitational/Streaming Instability (GI/SI).
The numerical simulations typically yield a size distribution that can be fit with an exponential
cutoff functional form \citep{2019ApJ...883..192A} that is also a good match to the
cold population. 
Here we have shown that it is also a good match to the hot population 
with $D \lesssim 300$~km.

Our paradigm is that after planetesimal formation, the conditions in the hot-population 
formation region  (in particular much higher surface density)
permitted the creation of bigger objects, which thus erased the taper for 
$H_r < 5.5$.
An important implication is that dwarf planets and larger objects then
accumulated mass without altering the relative size distribution in the 
$H_r \simeq 6$ to 9 range (and possibly much smaller sizes).
Two main mechanisms are usually invoked for this latter stage of accretion: 
runaway growth 
\citep[reviewed by][]{1993ARA&A..31..129L} 
and pebble accretion \citep[{\it e.g.},][]{2017ASSL..445..197O}.
Runaway growth does not care about the size of objects it sweeps up,
while in pebble accretion the very-big object mass accretion is from tiny pebbles.
In this later case, our result implies that objects with sizes corresponding to $H_r=$5.5--8.3
accumulate negligible further mass via pebbles.

If GI/SI is the planetesimal-formation mechanism,
scalings indicate that the mass of the largest bodies formed is governed by the local
solid surface density (to the third power) and heliocentric distance (to the 6th power)
\citep{2019ApJ...883..192A,2019ApJ...885...69L}.
For a surface density varying with heliocentric distance to the -1.5 or -2 power between 25 and 45~au, 
this would not be a problem as the density drop would roughly compensate for the increase in distance. However, there are arguments that the 
surface density drops by a factor of $\sim$1000 between these locations \citep[i.e.][]{2020AJ....160...46N}.
It is thus surprising that two populations would have the same exponential cutoff shape.
Some possible solutions to this dilemma are:

{\bf (1)}
    the initial planetesimal formation process is actually only mildly sensitive to the density, 
    but the creation of $H_r<5$ bodies in the hot-population formation region proceeded efficiently, while in
    the cold belt these dynamics were not triggered due to the low surface density,    or
    
{\bf (2)}
cold-population formation occurred in a localized over density 
(perhaps caused by a pressure bump) 
in an environment where the formed planetesimals are not confined to the overdense region,
while in contrast the hot population formed in an extended 
high-density zone where large-object  formation was 
possible through another process.

The first possibility is supported by \citet{2020ApJ...901...54K} who derive a criterion for the {\it minimum} object mass that can be created by GI/SI when accounting for diffusion due to turbulence. Based on this criterion, they claim that the size distribution should be a Gaussian centered on this minimum size ($D \simeq$80--85~km with width $\sim$45~km). Their minimum size is fairly insensitive to heliocentric distance from 3 to 30~au, but then drops markedly at larger distances.

The second possibility would have to be the solution if GI/SI is the dominant mechanism, unless the current theoretical scaling laws are incorrect.
In these studies, there is a trigger density one gets to and then GI/SI forms planetesimals quickly. 
This could result in the two regions having the same shape because they both reached the same critical density, 
although the way they achieved that density would have 
been different between the two populations.


Summarizing our discussion, we hypothesize that whatever mechanism created the first
planetesimals up to $\sim$400~km in diameter, it is largely insensitive to the global
physical conditions, and produces objects up to that size.
In dense environments, some other process(es) takes over to build bigger objects,
without altering the size distribution from $\sim$300~km down to a few 10s~km or even less.

\begin{acknowledgements}
{\bf Acknowledgements:}
\begin{footnotesize}
We thank Wes Fraser for useful discussions while preparing this manuscript,
and reviewer David Nesvorny for useful comments.
This work was supported by the Programme National de Plan\'etologie (PNP)
of CNRS-INSU co-funded by CNES.
BG ackowledges discovery grant funding support from NSERC.
This research made use of the Canadian Advanced Network for Astronomy
Research (CANFAR) and the facilities of the Canadian Astronomy Data
Centre operated by the National Research Council of Canada with the
support of the Canadian Space Agency.
This research has made use of data and/or services provided by the
International Astronomical Union's Minor Planet Center.
Based on observations obtained with MegaPrime/MegaCam,
a joint project of CFHT and CEA/DAPNIA, at the Canada-France-Hawaii
Telescope (CFHT) which is operated by the National Research Council
(NRC) of Canada, the Institut National des Sciences de l'Univers of
the Centre National de la Recherche Scientifique (CNRS) of France, and
the University of Hawaii. The observations at the CFHT were performed
with care and respect from the summit of Maunakea which is a
significant cultural and historic site.
\end{footnotesize}
\end{acknowledgements}

\bibliography{OtherPublications,Petit}
\bibliographystyle{aasjournal}

\appendix

\section{Kernel Density Estimator (KDE)}
\label{app:KDE}
To avoid the shortcomings of classical histograms due to size and location of
the bins, we use a Kernel Density Estimator (KDE)
\citep{rosenblatt1956,parzen1962} to estimate the true $H_r$ distribution (see
https://en.wikipedia.org/wiki/Kernel\_density\_estimation for an easy and basic
introduction to KDEs). The kernel is a non-negative function that smoothes the
contribution of each datum over an interval which size is determined by the
{\it bandwidth} parameter. There exists a variety of kernels; we use an
Epanechnikov finite extent kernel \citep{doi:10.1137/1114019}, as it is optimal
in a mean square error sense.

The bandwidth is an important parameter that
we determine using the empirical approach of cross-validation.
This empirical approach to model parameter selection does not depend on
(dubious) assumptions about the underlying data's distribution and thus 
is very flexible.
For the cold OSSOS++ sample the optimal bandwidth is 0.4, which we use to
plot Fig.~\ref{fig:KDE_H_OSSOS}.
For the hot sample, the optimal bandwidth depends rather strongly on the
inclusion or exclusion of the few objects at the tails of the distribution,
especially at the bright end where the $H_r$ spread is very uneven. This
means that the best bandwidth is very different at the large size and the
small size ends of the range, so we renormalize the $H_r$ values to make
them more uniform ($Y_r = 10^{0.1 (H_r-9)}$) and then use an optimal
bandwidth of 0.1 in this new variable.

In our computations, we use the implementation from the scikit-learn
python package \citep{scikit-learn}. We then scale the KDE to the
total number of objects in each component of the main classical
Kuiper Belt per magnitude and show the result in Fig.~\ref{fig:KDE_H_OSSOS}.

\section{Mass of the belt}
\label{app:Mass_belt}
We first derive the relation between $H_r$ and the TNO mass $M$.
The apparent magnitude $m_r$ in a given filter band (CFHTLS-$r'$ here)
is related to its radius $r$, in km, its geometric albedo $\nu_r$, 
its distance to the Sun $R$ and to the observer on Earth $\Delta$ (both in au), 
its phase angle $\gamma$, the phase function $\Phi(\gamma)$, the rotational 
lightcurve function $f(t)$ and the magnitude of
the Sun $m_{r,Sun}$ in the same band, by
\begin{equation}
  \label{eq:m_msun}
m_r = m_{r,Sun} - 2.5 \log_{10}\left(\nu_r \left(\frac{r}{1{\rm ~km}}\right)^2 \Phi(\gamma) f(t)\right) + 2.5 \log_{10}\left(2.25 \times 10^{16} \frac{R^2 \Delta^2}{{\rm au}^4}\right)
\end{equation}
Here we assume $f(t) = 1$ as the average of a rapid periodic function.
From $H_r$ definition, we have
\begin{eqnarray*}
H_r & = & m_r + 2.5 \log_{10}(\Phi(\gamma)) - 2.5 \log_{10}\left(\frac{R^2 \Delta^2}{{\rm au}^4}\right)) \\
    & = & m_{r,Sun} + 2.5 \log_{10}\left(2.25 \times 10^{16}\right) - 2.5 \log_{10}\left(\nu_r \left(\frac{r}{1{\rm ~km}}\right)^2\right)
\end{eqnarray*}

For the cfhtls-r filter, $m_{r,Sun} = -26.94$ in the AB system 
\citep{sunAB} and  therefore
\begin{equation}
\label{eq:final_H_r}
H_r = 13.94 - 2.5 \log_{10}\left(\nu_r \left(\frac{r}{1{\rm ~km}}\right)^2\right)
\end{equation}

Denoting bulk density as $\rho$ , their mass
  $M = \frac{4 \pi}{3} \rho r^3$
  can be written, using
Eq.~\ref{eq:final_H_r}, as
\begin{equation}
\label{eq:density_H_r_nu}
M = \frac{4 \pi}{3} \frac{\rho}{\nu_r^{3/2}} 10^{(13.94 - H_r)\frac{3}{5}} = B \frac{\rho}{\nu_r^{3/2}} 10^{-\frac{3 H_r}{5}}
\end{equation}
with 
$B = \frac{4 \pi}{3} 10^{13.96\frac{3}{5}}$~kg~$= 9.685 \times 10^8$~kg, and $\rho$ given in kg.km$^{-3}$ (1~g.cm$^{-3}$=~10$^{12}$~kg.km$^{-3}$).

The computation of TNO mass from its absolute magnitude $H_r$ thus strongly
depends on the two unknowns $\rho$ and $\nu_r$. Conservatively assuming that
$\rho$ can vary from 0.5~g.cm$^{-3}$ to 2~g.cm$^{-3}$, and $\nu_r$ from 0.06 (as
seen from the comets), to 0.24 (Arrokoth), we can formally have a variation by
a factor of $\sim$30 in the mass of individual objects.

\subsection{The hot belt}
\label{app:Mass_hot}
For a scenario of fixed values of the albedo and the bulk density, the mass 
of the hot belt is dominated by the few largest bodies (such as 
Makemake, Quaoar, Varda or Varuna) but 
the predicted mass of these large bodies is badly overestimated if using
nominal values of these quantities.
With the nominal values of $\nu_r = 0.08$ \citep{2014ApJ...793L...2L}, $\rho = 1$g/cm$^3$ \citep{2001AJ....122.1051G,2014ApJ...782..100F} and $H_r = -0.31$, one would find a mass for Makemake of $1.1 \; 10^{-2} M_\Earth$ when its actual mass is $5.2 \; 10^{-4} M_\Earth$ \citep{2018DPS....5050902P}. 
The discrepancy decreases with increasing $H_r$ but is still of a factor of 2 for Varuna at $H_r = 3.59$.

The simple solution of this issues is that
\citet{2008ssbn.book..161S} showed a correlation between
object size and geometric albedo, which \citet{2008Icar..195..827F} modelled as $\nu \propto r^\beta$. 
It is also likely that the density of a body increases with its size due to self-compression. 
In the mass-to-$H_r$ relation, the important factor is $\rho \nu^{-3/2}$ which 
we then model as $\rho \nu^{-3/2} = A (M/1{\rm ~kg})^\gamma$. 
We select $\gamma=0.77$ with resulting 
$A \sim 6.7 \times 10^{28}$ for $\rho$ expressed in kg.km$^{-3}$   
to roughly match the known masses of Makemake, Quaoar ($M \simeq 1.4 \; 10^{21}$kg,
\citet{2012A&A...543A..68V,2013Icar..222..357F}), Varda ($M \simeq 2.45 \;
10^{20}$kg, \citet{2020A&A...643A.125S}), Varuna ($1.5 \; 10^{20}$kg,
\citet{2007AJ....133.1393L,2013A&A...557A..60L}) and (55637) 2002~UX25 ($1.2 \;
10^{20}$kg, \citet{2013ApJ...778L..34B}). We use this dependency for the large
bodies until the $\rho \nu^{-3/2}$ factor reaches its nominal value for $\rho =
1$g/cm$^3$ and $\nu_r = 0.08$, which occurs at $M = 5.6 \; 10^{19}$kg and $H_r =
4.82$.  For smaller masses (larger $H_r$), we use the nominal values for $\rho$
and $\nu$.


Brightward of $H_r = 5.47$, we use the raw MPC absolute magnitude distribution.
Between $H_r = 5.47$ and $H_r = 8.3$ we use the debiased $H_r$ distributions
shown in Fig.~\ref{fig:KDE_H_OSSOS}. Faintward of this limit, we use the
exponential cutoff from the cold belt (eq.~\ref{eqn:LF_cold}) scaled-up by a factor of 2.2 as in Figure~\ref{fig:KDE_H_shift}. As expected the mass is not concentrated at any given size. The total mass of the hot main classical belt is 0.012~$M_\Earth$.

\subsection{The cold belt}
\label{app:Mass_cold}
For the cold population, all TNOs are small enough to be in the regime where 
albedos and densities that are not correlated with their size. 
We use the nominal $\rho = 1000$kg/m$^3$ with more reflective $\nu_r = 0.15$
\citep{2014ApJ...782..100F,2014ApJ...793L...2L} estimated for cold
objects. 
The debiased mass of the cold
belt, brighter than $H_r = 8.3$ comes out as 0.0010~$M_\Earth$.

For cold TNOs with $H_r>8.3$, we use the exponential cutoff parameterization 
from \citet{2021ApJ...920L..28K},
\begin{equation}\label{eqn:LF_cold}
    N(<H_r)=  10^{\alpha(H_r-H_o)} \; \exp\left[ -10^{-\beta (H_r-H_B)}\right]
\end{equation}
where $N$ is the total population for TNOs with absolute magnitude less than $H_r$;
$H_o = -2.6$ is a normalization factor, 
$\alpha = 0.4$ is the asymptotic logarithmic slope at large $H_r$, 
$\beta = 0.25$ is the strength of the exponential tapering and 
$H_B = 8.1$ is the $H_r$ value at which the exponential taper begins to dominate
as one moves towards brighter magnitudes\footnote{Note that numbers in Table~\ref{tab:pop_estimate} are obtained from debiasing of OSSOS data, not from this formula.}.
We caution that there is degeneracy in this parameterization, which allows 
individual parameters to have large variations as long as the others change
in a correlated way, which results in the cumulative population curve being
very similar.
This yields a mass of 0.0011~$M_\Earth$ for cold $H_r>8.3$~TNOs, assuming 
$\alpha=0.4$ continues.

Thus the total mass of the cold belt is 0.0021~$M_\Earth$, for the assumed albedo and bulk density, or about 1/6 of our hot-belt mass estimate.

\listofchanges

\end{document}